\documentclass[10pt, conference, compsocconf]{IEEEtran}
\IEEEoverridecommandlockouts
\usepackage{acronym}
\usepackage{booktabs}
\usepackage{cite}
\ifCLASSINFOpdf
  \usepackage[pdftex]{graphicx}
\else
  \usepackage[dvips]{graphicx}
\fi
\usepackage[cmex10]{amsmath}
\usepackage{algorithm}
\usepackage[noend]{algpseudocode}
\usepackage[tight,footnotesize]{subfigure}
\usepackage{url}
\usepackage{color}

\acrodef{BF}{Brute Force}
\acrodef{GB}{Grid Based}
\acrodef{DDM}{Data Distribution Management}
\acrodef{ITM}{Interval Tree Matching}
\acrodef{SBM}{Sort-Based Matching}
\acrodef{BF}{Brute Force}
\acrodef{MMOG}{Massive Multiplayer Online Game}
\acrodef{HLA}{High Level Architecture}

\begin{document}
\title{A Parallel Data Distribution Management Algorithm\footnotemark}

\author{\IEEEauthorblockN{Moreno Marzolla, Gabriele D'Angelo}
\IEEEauthorblockA{Dept. of Computer Science and Engineering\\
University of Bologna, Italy\\
Email: moreno.marzolla@unibo.it, g.dangelo@unibo.it}
\and
\IEEEauthorblockN{Marco Mandrioli}
\IEEEauthorblockA{Email: m.mandrioli@live.it}
}

\maketitle

\footnotetext{The publisher version of this paper is available at \url{http://dx.doi.org/10.1109/DS-RT.2013.23}. \textbf{{\color{red}Please cite as: Moreno Marzolla, Gabriele D'Angelo, Marco Mandrioli. A Parallel Data Distribution Management Algorithm. Proceedings of the 2013 IEEE/ACM 17th International Symposium on Distributed Simulation and Real Time Applications (DS-RT 2013). ISBN, 978-0-7695-5138-8.}}}

\begin{abstract}
Identifying intersections among a set of $d$-dimensional rectangular
regions ($d$-rectangles) is a common problem in many simulation and
modeling applications. Since algorithms for computing intersections
over a large number of regions can be computationally demanding, an
obvious solution is to take advantage of the multiprocessing
capabilities of modern multicore processors. Unfortunately, many
solutions employed for the Data Distribution Management service of the
High Level Architecture are either inefficient, or can only partially
be parallelized. In this paper we propose the Interval Tree Matching
(ITM) algorithm for computing intersections among $d$-rectangles. ITM
is based on a simple Interval Tree data structure, and exhibits an
embarrassingly parallel structure. We implement the ITM algorithm, and
compare its sequential performance with two widely used solutions
(brute force and sort-based matching). We also analyze the scalability
of ITM on shared-memory multicore processors. The results show that
the sequential implementation of ITM is competitive with sort-based
matching; moreover, the parallel implementation provides good speedup
on multicore processors.
\end{abstract}


\section{Introduction}\label{sec:introduction}

The~\ac{HLA} specification~\cite{HLA} defines several~\ac{DDM}
services to forward events generated on \emph{update} regions to a set
of \emph{subscription} regions. For example, consider a simulation of
vehicles moving on a two-dimensional terrain. Each vehicle may be
interested in events happening inside its area of interest (e.g., its
field of view) that might be approximated with a rectangular region
centered at the vehicle position. This kind of problem also arises in
the context of Massively Multiplayer Online Games, where the game
engine must send updates only to players that might be affected by
game events, in order to reduce computation cost and network
traffic. In this paper we assume that a region corresponds to a single
\emph{extent} in~\ac{DDM} terminology), that is, a $d$-dimensional
rectangle ($d$-rectangle) in a $d$-dimensional routing space.

Spatial data structures that can solve the region intersection problem
have been developed over the years; examples include the $k$-$d$
tree~\cite{Rosenberg1985} and R-tree~\cite{Guttman1984}. However, it
turns out that simpler, less efficient solutions are actually
preferred in practice and widely deployed in~\ac{DDM}
implementations. The reason is that efficient spatial data structures
tend to be complex to implement, and therefore their theoretical
performance is affected by high constant factors.

The increasingly large size of computer simulations employing~\ac{DDM}
techniques is posing a challenge to the existing solutions. As the
number of regions increases, so does the execution time of
the~\ac{DDM} service. Given the current trend in microprocessor design
where a single CPU contains multiple independent execution units,
significant improvements could be achieved if the existing~\ac{DDM}
matching algorithms were capable of taking advantage of the
computational power provided by multi-core processors.

There are two opportunities for parallelizing~\ac{DDM} algorithms.
The first is based on the observation that the problem of identifying
whether two $d$-rectangles intersect can be reduced to $d$ independent
intersection problems among one-dimensional segments (details will be
given in Section~\ref{sec:ddm-algorithms}). Therefore, given an
algorithm that can identify intersections among two sets of segments,
we can execute $d$ instances in parallel, each computing the
intersections among the projections of the extents along each
dimension. The extent intersections can be easily computed from the
segments overlap information.

The idea above can be regarded as the ``low hanging fruit'' which is
easy to get, but does not solve the problem in the long run.  In fact,
the number of cores in modern processors is often larger than the
number of dimensions of most routing spaces; this gap is likely to
increase (e.g., the Tilera TILE-Gx8072 processor~\cite{tileragx}
offers 72 general-purpose cores on the same chip, connected through an
on-chip mesh network). Here comes the second parallelization
opportunity: distribute the regions to the available cores so that
each core can work on a smaller problem. This is quite difficult to
achieve on the existing~\ac{DDM} algorithms, since they are either
inefficient (and therefore there is little incentive in splitting the
workload), or inherently sequential (and therefore there is no easy
way to achieve parallelism over the set of extents).

In this paper we describe the~\acf{ITM} algorithm for solving the
one-dimensional segment intersection problem. The algorithm uses a
simple implementation of the Interval Tree data structure based on an
augmented balanced search tree. Experimental performance measures
indicate that the sequential version of~\ac{ITM} is competitive in the
sequential case with the best algorithm used for~\ac{DDM}, namely
sort-based matching. We also observed good scalability of the parallel
implementation of~\ac{ITM} on shared-memory architectures. An
important feature of~\ac{ITM} is that it can be used to efficiently
update overlap information in a dynamic setting, that is, in case
extents can be moved or resized dynamically.

This paper is organized as follows. In Section~\ref{sec:related-work}
we briefly review the state of the art and compare~\ac{ITM} with
existing solutions to the~\ac{DDM} matching problem. In
Section~\ref{sec:ddm-algorithms} we describe three commonly used
algorithms for~\ac{DDM}: brute force, grid-based and sort-based
matching. In Section~\ref{sec:parallel-ddm} we describe~\ac{ITM} and
analyze its computational cost. In
Section~\ref{sec:experimental-evaluation} we experimentally evaluate
the performance of the sequential version of~\ac{ITM} compared with
brute force and sort-based matching; additionally, we study the
scalability of a parallel implementation of~\ac{ITM} on a multicore
processor.  Finally, conclusions and future works will be discussed in
Section~\ref{sec:conclusions}.

\section{Related Work}\label{sec:related-work}

\ac{DDM} matching can be considered as an instance of the more general
problem of identifying intersecting pairs of (hyper-)rectangles in a
multidimensional metric space. Well known space-partitioning data
structures such as $k$-$d$ trees~\cite{Rosenberg1985} and
R-trees~\cite{Guttman1984} can be used to efficiently store volumetric
objects and identify intersections with a query object. However,
spatial data structures are quite complex to implement and, although
asymptotically efficient, they can be slower than less efficient but
simpler solutions in many real-world situations~\cite{petty-1997}.
In~\cite{Devai2010} the authors describe a rectangle-intersection
algorithm in two-dimensional space that uses only simple data
structures (arrays), and can enumerate all $k$ intersections among $n$
rectangles $O(n \log n + k)$ time and $O(n)$ space.

The usage of Interval Trees for~\ac{DDM} was first proposed
in~\cite{petty-1997} where the authors used a different and more
complex data structure than the one proposed here (see
Section~\ref{sec:parallel-ddm}). In their case, the performance
evaluation on very small instances shows mixed results.

\ac{SBM}~\cite{Raczy2005} is a widely used algorithm for enumerating
all intersections among subscription and update extents, with
particular emphasis on distributed simulation applications based on
the High Level Architecture (HLA) specification. \ac{SBM} first sorts
the endpoints, and then scans the sorted set (details will be given in
Section~\ref{sec:sort-based}). \ac{SBM} is extended in~\cite{Pan2011}
to work efficiently on a dynamic scenario where extents can be moved
or resized dynamically.

Despite its simplicity and efficiency, \ac{SBM} has the drawback that
its sequential scan step is intrinsically serial and can not be easily
parallelized. This can be a serious limitation when dealing with a
large number of extents on multicore processors.

In~\cite{6147978} the authors propose a binary partition-based
matching algorithm that has good performances in some settings, but
suffers from a worst case cost of $O(N^2\log N)$ where $N$ is the
total number of subscription and update regions. Moreover, the
extension of this algorithm to the dynamic scenario seems impractical.

Layer et al.~\cite{Layer2012} describe the Binary Interval Search
(BITS) algorithm. BITS can be used to efficiently count the number of
intersections between two sets $A$ and $B$ of intervals in time
$O\left( (|A| + |B|) \log |B| \right)$. To do so, BITS performs a
preprocessing phase in which two sorted arrays $B_S$ and $B_E$ are
created in time $O(|B| \log |B|)$. $B_S$ contains the starting points
of all intervals in $B$, while $B_E$ contains the ending points. The
number of intervals in $B$ that intersect a given query interval $q =
[q.\textit{low}, q.\textit{high}]$ can be computed by subtracting from
$|B|$ the number of intervals which do \emph{not} intersect $q$. BITS
uses two binary searches in $B_S$ and $B_E$ to compute the number of
intervals in $B$ whose ending point precedes $q.\textit{low}$, and
those whose starting point follows $q.\textit{high}$. While BITS can
be easily parallelized by executing the binary searches in parallel,
it must be observed that the problem of enumerating all intersections
can not be easily handled by BITS without substantial modifications
which significantly increase its computational cost.

\section{DDM Matching Algorithms}\label{sec:ddm-algorithms}

In this section we define the~\ac{DDM} problem and describe three well
known solution algorithms that have been thoroughly investigated.

Let $\mathbf{S} = \{S_1, \ldots, S_n\}$ and $\mathbf{U}=\{U_1, \ldots,
U_m\}$ be two sets of rectangular regions in $d$-dimensional space
($d$-rectangles, also called \emph{extents}). $\mathbf{S}$ is the set
of \emph{subscription extents}, while $\mathbf{U}$ is the set of
\emph{update extents}. Each extent $T$ has an integer attribute
$T.\textit{id}$ representing its index in the set it belongs to, e.g.,
$S_i .\textit{id} = i$ and $U_j .\textit{id} = j$. The goal of
a~\ac{DDM} matching algorithm is to identify all intersections between
a subscription and an update extent, that is, enumerating the content
of the subset of $\mathbf{S} \times \mathbf{U}$ defined as
\[
\{S_i \in \mathbf{S},\ U_j \in \mathbf{U}\ |\ S_i \cap U_j \neq \emptyset\}
\]

\begin{figure}[t]
\centering\includegraphics[scale=.7]{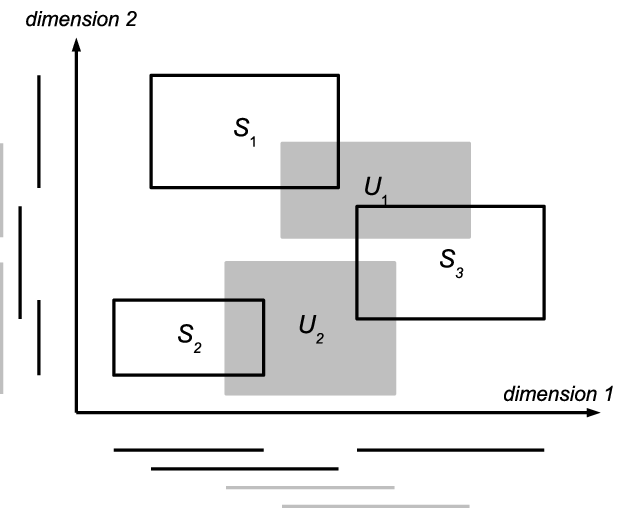}
\caption{Data Distribution Management example in $d=2$ dimensions}\label{fig:ddm_example}
\end{figure}

Figure~\ref{fig:ddm_example} shows an example in $d=2$ dimensions with
three subscription extents $\{S_1, S_2, S_3\}$ and two update extents
$\{U_1, U_2\}$. We observe that $U_1$ overlaps with $S_1$ and $S_3$,
while $U_2$ overlaps with $S_2$ and $S_3$.

The preferred way of storing the intersections uses a $n \times m$
binary matrix $\mathbf{M}$, where each element $M_{ij} = 1$ if and
only if $S_i$ intersects $U_j$. In the case of
Figure~\ref{fig:ddm_example} we have
\[
\mathbf{M} = \left(\begin{array}{cc}
1 & 0 \\
0 & 1 \\
1 & 1 
\end{array}\right)
\]
Since the number of intersections is generally much smaller than $n
\times m$, matrix $\mathbf{M}$ tends to be sparse and can be stored in
compressed form to reduce the memory requirement.

It is important to observe that any~\ac{DDM} algorithm that enumerates
all $K$ intersections requires time at least $\Omega(K)$; we say that
the time complexity of~\ac{DDM} matching algorithms is
\emph{output-sensitive}, since it depends on the size of the output as
well as on the size of the input. Since $K \leq nm$, in the worst case
we have that any algorithm has a worst-case complexity of
$O(nm)$. Every algorithm that stores the result into an uncompressed
intersection matrix requires time $O(nm)$ to initialize the matrix,
regardless of the number of intersections. Despite this, it makes
sense to try to improve the efficiency of overlap identification,
since in practice this is the slower step of the~\ac{DDM} problem.

\subsection{Testing Intersection}\label{sec:testing-intersection}

\begin{algorithm}[t]
\caption{Segment intersection test}\label{alg:intersect1d}
\begin{algorithmic}
\Function{Intersect-1D}{$x, y$}
\State \textbf{return} $x.\emph{low} < y.\emph{high} \wedge y.\emph{low} < x.\emph{high}$
\EndFunction
\end{algorithmic}
\end{algorithm}

Testing whether two $d$-rectangles intersect is a key operation. When
$d=1$ the problem is reduced to testing whether two segments
$x=[x.\textit{low},x.\textit{high}]$, $y=[y.\textit{low},
  y.\textit{high}]$ intersect, that can be done in time $O(1)$ using
Algorithm~\ref{alg:intersect1d}.

For the general case $d>1$ we observe that two $d$-dimensional extents
$S_i$ and $U_j$ intersect if and only if all their projections along
each dimension intersect. Looking again at
Figure~\ref{fig:ddm_example}, we see that the projections of $S_1$ and
$U_2$ intersect along dimension 1 but not along dimension 2;
therefore, $S_1$ and $U_2$ cannot intersect. On the other hand, the
projections of $S_2$ and $U_2$ intersect along both dimensions, and in
fact these rectangular regions intersect in the plane.

Since dealing with segments is easier than dealing with
$d$-rectangles, it is common practice in the~\ac{DDM} research
community to define efficient algorithms for the one-dimensional case,
and use them to solve the general higher dimensional case. According
to the discussion above, an algorithm that enumerates all
intersections among two sets of $n$ and $m$ segments in time
$O\left(f(n,m)\right)$ can be immediately extended to an $O\left( d
\times f(n,m) \right)$ algorithm for $d$-rectangles.  For this reason
in the rest of this paper we will consider the case $d=1$ only.

\subsection{Region-Based Matching}

\begin{algorithm}[t]
\caption{Brute Force Matching (BF)}\label{alg:brute-force}
\begin{algorithmic}
\Function{BruteForce-1D}{$\mathbf{S}, \mathbf{U}$}
\State $n \gets |\mathbf{S}|$, $m \gets |\mathbf{U}|$
\State Let $\mathbf{M}$ be an $n \times m$ intersection matrix
\ForAll{$i \gets 1, n$}
\ForAll{$j \gets 1, m$}
\State $M_{ij} \gets$ \Call{Intersect-1D}{$S_i, U_j$}
\EndFor
\EndFor
\State \textbf{return} $\mathbf{M}$
\EndFunction
\end{algorithmic}
\end{algorithm}

The most direct approach for solving the segment intersection problem
is Region-Based matching, also called~\acf{BF} approach shown in
Algorithm~\ref{alg:brute-force}. The~\ac{BF} algorithm tests all $n
\times m$ subscription-update pairs, and records intersection
information in matrix $\mathbf{M}$.

The~\ac{BF} algorithm requires time $O(nm)$, and is therefore not very
efficient; despite this, it is appealing due to its
simplicity. Furthermore, \ac{BF} can be trivially parallelized since
all iterations are independent (it is an example of
\emph{embarrassingly parallel} computation). When $p$ processors are
available, the amount of work performed by each processor is
$O\left(nm / p\right)$.

\subsection{Grid-Based Matching}\label{sec:grid-based-matching}

\begin{figure}[t]
\centering\includegraphics[scale=.7]{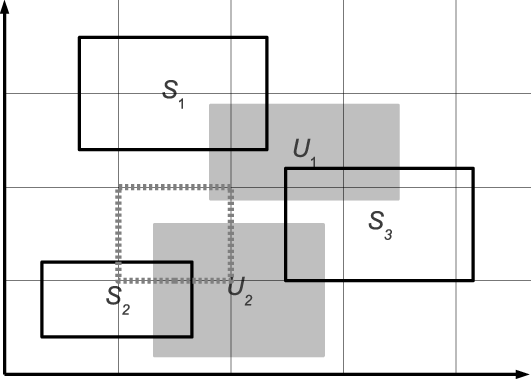}
\caption{Grid-based DDM in $d=2$ dimensions.}\label{fig:ddm_example_grid}
\end{figure}

The~\acf{GB} matching algorithm proposed by Boukerche and
Dzermajko~\cite{Boukerche2001} is an improved solution to the
$d$-rectangle intersection problem. It works by partitioning the
routing space into a grid of $d$-dimensional cells. Each extent is
mapped to the grid cells it overlaps with. The events produced by an
update extent $U_j$ are sent to all subscriptions that share at least
one cell in common with $U_j$.

The~\ac{GB} approach is more scalable than~\ac{BF}; furthermore, its
performance can be tuned by choosing a suitable cell
size. Unfortunately, it has some drawbacks: \ac{GB} matching may
report spurious overlaps, that is, may deliver events to subscribers
which should not receive them. This situation is illustrated in
Figure~\ref{fig:ddm_example_grid}: the extent $U_1$ and $S_2$ share
the dashed cell but do not overlap; therefore, $S_2$ will receive
spurious notifications from $U_1$ that will need to be filtered out at
the receiving side.

The problem of spurious events can be mitigated by applying the brute
force algorithm to each grid cell. If the routing space is partitioned
into $G$ cells and all extents are evenly distributed over the grid,
each cell will have $n/G$ subscription extents and $m/G$ update
extents. Therefore, the brute force approach applied to each cell
requires $O(nm / G^2)$ operations; since there are $G$ cells, the
overall complexity becomes $O(nm / G)$. In conclusion, in the ideal
case the~\ac{GB} matching can reduce the workload by a factor $G$ with
respect to~\ac{BF}. Unfortunately, when cells are small (and therefore
$G$ is large) each extent is mapped to a larger number of cells, which
increases the computation time.

\subsection{Sort-Based Matching}\label{sec:sort-based}

The~\acl{SBM} algorithm proposed by Raczy et al.~\cite{Raczy2005} is a
simple and very efficient solution to the~\ac{DDM} matching
problem. 

\begin{algorithm}[t]
\caption{Sort-Based Matching (SBM)}\label{alg:sort-based}
\begin{algorithmic}
\Function{Sort-Based-Matching-1D}{$\mathbf{S}, \mathbf{U}$}
\State $n \gets |\mathbf{S}|$, $m \gets |\mathbf{U}|$
\State Let $\mathbf{M}$ be an $n \times m$ intersection matrix
\State Let $L$ be a vector with $2(n+m)$ elements
\ForAll{extents $x \in \mathbf{S} \cup \mathbf{U}$}
\State Insert $x.\textit{lower}$ and $x.\textit{upper}$ in $L$
\EndFor
\State Sort $L$ in nondecreasing order
\State $\textit{SubscriptionSet} \gets \textit{UpdateSet} \gets \emptyset$
\ForAll{points $p \in L$ in nondecreasing order}\label{alg:sbm-loop}
\If{$p$ belongs to subscription extent $T$}
\If{$p$ is the lower bound of $T$}
\State $\textit{SubscriptionSet} \gets \textit{SubscriptionSet} \cup \{T\}$
\Else
\State $\textit{SubscriptionSet} \gets \textit{SubscriptionSet} \setminus \{T\}$
\State\Comment extents in $\textit{UpdateSet}$ overlap $T$
\State $i \gets T.\textit{id}$
\ForAll{$x \in \textit{UpdateSet}$}
\State $j \gets x.\textit{id}$
\State $\mathbf{M}_{ij} \gets 1$
\EndFor
\EndIf
\Else\Comment{$T$ is an update extent}
\If{$p$ is the lower bound of $T$}
\State $\textit{UpdateSet} \gets \textit{UpdateSet} \cup \{T\}$
\Else
\State $\textit{UpdateSet} \gets \textit{UpdateSet} \setminus \{T\}$
\State\Comment extents in $\textit{SubscriptionSet}$ overlap $T$
\State $j \gets T.\textit{id}$
\ForAll{$x \in \textit{SubscriptionSet}$}
\State $i \gets x.\textit{id}$
\State $\mathbf{M}_{ij} \gets 1$
\EndFor
\EndIf
\EndIf
\EndFor
\State \textbf{return} $\mathbf{M}$
\EndFunction
\end{algorithmic}
\end{algorithm}

In its basic version, \ac{SBM} is illustrated in
Algorithm~\ref{alg:sort-based}. Given a set $\mathbf{S}$ of $n$
subscription intervals, and a set $\mathbf{U}$ of $m$ update
intervals, the algorithm sorts the endpoints in nondecreasing order in
the array $L$. Then, the algorithm performs a scan of the sorted
vector; two sets $\textit{SubscriptionSet}$ and $\textit{UpdateSet}$
are used to keep track of the active subscription and update intervals
at every point $p$.  Each time the upper bound of an interval $T$ is
encountered, the intersection matrix $\mathbf{M}$ is updated
appropriately, depending on whether $T$ is a subscription or update
extent.

As can be seen, \ac{SBM} uses only simple data structures. If we
ignore the time needed to initialize the matrix $\mathbf{M}$,
Algorithm~\ref{alg:sort-based} requires time $O\left( (n+m) \log (n+m)
\right)$ to sort the vector $L$, then time $O(n+m)$ to scan the sorted
vector. During the scan phase, total time $O(nm)$ is spent to transfer
the information from the sets \textit{SubscriptionSet} and
\textit{UpdateSet} to the intersection matrix $\mathbf{M}$, assuming
that the sets above are implemented as bitmaps~\cite{Raczy2005}. The
overall computational cost is $O\left( (n+m) \log (n+m) + nm \right)$,
and therefore asymptotically not better than~\ac{BF}; however, the
term $O(nm)$ comes from simple operations on bitmaps, hence~\ac{SBM}
is very efficient in practice~\cite{Raczy2005}.

While~\ac{SBM} is very fast, it has the drawback of not being easily
parallelizable. In fact, while parallel algorithms for sorting the
array $L$ are known~\cite{parallel-merge-sort}, the scan step is
affected by loop-carried dependencies, since the content of
\textit{SubscriptionSet} and \textit{UpdateSet} depend on their
values at the previous iteration. This dependency can not be easily
removed. Given the widespread availability of multi- and many-core
processors, this limitation can not be ignored.

In the next section we introduce the~\acl{ITM} algorithm for computing
intersections among two sets of intervals. \ac{ITM} uses an augmented
AVL tree data structure to store the intervals.  The performance
of~\ac{ITM} depends on the number of intersections; however we will
show that~\ac{ITM} is faster than~\ac{SBM} in the scenarios considered
in the literature.  Furthermore, \ac{ITM} can be trivially
parallelized, hence further performance improvements can be obtained
on shared-memory multi-core processors.

\section{Interval Tree Matching}\label{sec:parallel-ddm}

\ac{ITM} is a~\ac{DDM} matching algorithm for one dimensional segments
based on the \emph{Interval Tree} data structure. An Interval Tree
stores a dynamic set of $n$ intervals, and supports insertions,
deletions, and queries to get the list of segments intersecting with a
given interval $q$. 

Different implementations of the Interval Tree are possible. Priority
search trees~\cite{McCreight85} support insertions and deletions in
time $O(\log n)$, and can report all $k$ intersections with a given
query interval in time $O(k + \log n)$.  For the experimental
evaluation described in Section~\ref{sec:experimental-evaluation} we
implemented the simpler but less efficient variant based on augmented
AVL trees~\cite{avl}, described in~\cite[Chapter 14.3]{Cormen2009}. We
did so in order to trade a slight decrease in asymptotic efficiency
for a simpler and more familiar data structure. It should be observed
that~\ac{ITM} is not tied to any specific implementation of Interval
Tree, therefore any data structure can be used as a drop-in
replacement inside the algorithm.

Each node $x$ of the AVL tree holds an interval $x.\textit{in}$;
intervals are sorted according to their lower bounds, and ties are
broken by comparing upper bounds. Node $x$ includes two additional
fields $x.\textit{maxupper}$ and $x.\textit{minlower}$, representing
the maximum value of the upper bound and minimum value of the lower
bound, respectively, of all intervals stored in the subtree rooted at
$x$. We have chosen AVL trees over other balanced search trees, such
as red-black trees~\cite{rbtree}, because AVL trees are more rigidly
balanced and therefore allow faster queries.

\begin{figure}[t]
\centering\includegraphics[width=\columnwidth]{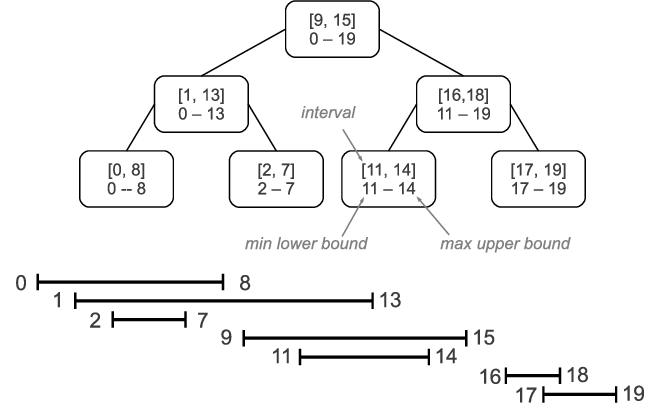}
\caption{Interval Tree representation of a set of intervals}\label{fig:interval_tree}
\end{figure}

Figure~\ref{fig:interval_tree} shows an example of Interval Tree with
$n=7$ intervals. Insertions and deletions are handled with the usual
rules of AVL trees, with the additional requirement to propagate
updates of the $\textit{maxupper}$ and $\textit{minlower}$ attributes
up to the root. Since the height of an AVL tree is $O(\log n)$,
insertions and deletions in the augmented data structure still require
$O(\log n)$ time in the worst case. The storage requirement is $O(n)$.

\begin{algorithm}[t]
\caption{Returns the list of intervals intersecting $q$}\label{alg:query}
\begin{algorithmic}
\Function{Interval-Query}{$x, q, \mathbf{M}$}

\If{$x = \textit{null}$ \textbf{or} $x.\textit{maxupper} < q.\textit{lower}$ \textbf{or} $x.\textit{minlower} > q.\textit{upper}$}
\State \textbf{return}
\EndIf

\State \Call{Interval-Query}{$x.\textit{left}, q, \mathbf{M}$}

\If{\Call{Intersect-1D}{$x.\textit{in}, q$}}
\State $i \gets x.\textit{in}.\textit{id}$, $j \gets q.\textit{id}$
\State $\mathbf{M}_{ij} \gets 1$
\EndIf

\If{$q.\textit{upper} > x.\textit{in}.\textit{lower}$}
\State \Call{Interval-Query}{$x.\textit{right}, q, \mathbf{M}$}
\EndIf

\EndFunction
\end{algorithmic}
\end{algorithm}

Function $\textsc{Interval-Query}(x, q, \mathbf{M})$, described in
Algorithm~\ref{alg:query}, is used to update matrix $\mathbf{M}$ with
all intersections of the update extent $q$ with the segments stored in
the subtree rooted at node $x$. The function is invoked as
$\textsc{Interval-Query}(T.\textit{root}, q, \mathbf{M})$. The basic
idea is very similar to a conventional item lookup in a binary search
tree, the difference being that at each node $x$ both the
$x.\textit{minlower}$ and $x.\textit{maxupper}$ fields are used to
drive the exploration of the tree; also, the search might proceed on
both the left and right child of node $x$. Algorithm~\ref{alg:query}
can identify all $k$ intersections between $q$ and all $n$ intervals
stored in the tree $T$ in time $O\left(\min\{n, (k+1) \log
n\}\right)$.

\begin{algorithm}[t]
\caption{Interval Tree Matching Algorithm}\label{alg:itm}
\begin{algorithmic}
\Function{\textsc{IntTree-Matching-1D}}{$\mathbf{S}, \mathbf{U}$}
\State $n \gets |\mathbf{S}|$, $m \gets |\mathbf{U}|$
\State Let $\mathbf{M}$ be an $n \times m$ intersection matrix
\State $T \gets \textsc{Interval-Tree-Create}(\mathbf{S})$
\ForAll{$j \gets 1, m$}
\State \Call{Interval-Query}{$T.\textit{root}, U_j , \mathbf{M}$}
\EndFor
\State \textbf{return} $\mathbf{M}$
\EndFunction
\end{algorithmic}
\end{algorithm}

The complete~\ac{ITM} matching procedure can now be easily described
in Algorithm~\ref{alg:itm}. First, an Interval Tree $T$ is created
from the subscription extents in $\mathbf{S}$.  Then, for each update
extent $U_j \in \mathbf{U}$, function \textsc{Interval-Query} is
invoked to identify all subscriptions that intersect $U_j$.

\paragraph*{Asymptotic Running Time} If there are
$n$ subscription and $m$ update extents, the Interval Tree of
subscriptions can be created in time $O(n \log n)$ and requires space
$O(n)$; the total query time is $O\left( \min\{mn, (K+1) \log
n\}\right)$, $K \leq nm$ being the number of intersections involving
all subscription and all update intervals. Note that we can assume
without loss of generality that $n \leq m$ (if this is not the case,
we can switch the role of $\mathbf{S}$ and $\mathbf{U}$).

\paragraph*{Parallelizing~\ac{ITM}}
Algorithm~\ref{alg:itm} can be trivially parallelized, since all $m$
queries on $T$ are independent. Note that function
\textsc{Interval-Query} modifies the intersection matrix $\mathbf{M}$
passed as parameter; however, each invocation of
\textsc{Interval-Query} modifies a different column of $\mathbf{M}$,
therefore no conflicts arise.  In
Section~\ref{sec:experimental-evaluation} we will illustrate the
results of experimental investigations on the scalability of the
parallel implementation of~\ac{ITM}.

\paragraph*{Dynamic interval management} 
Another interesting feature of~\ac{ITM} is that it can easily cope
with \emph{dynamic} intervals. In most applications, extents can move
and grow/shrink dynamically; if an update extent, say $U_j$, changes
its position or size, then it is necessary to recompute column $j$ of
matrix $\mathbf{M}$. The brute force approach applied to $U_j$ alone
gives an $O(n)$ algorithm, since it is only necessary to identify
overlaps between $U_j$ and all $n$ subscription segments. An extension
of~\ac{SBM} capable of updating intersection information efficiently
has been proposed~\cite{Pan2011}, with an asymptotic cost that depends
on various factors (e.g., upper bound of the dimension, maximum bound
shift in a region modification). On the other hand, we can use two
Interval Trees $T_U$ and $T_S$, holding the set of update and
subscription extents, respectively, to recompute the intersections
efficiently. If an update extent $U_j$ is modified, we can identify
the subscriptions overlapping $U_j$ in time $O\left( \min \{n,
(k+1)\log n \}\right)$ by performing a query on $T_S$. Similarly, if a
subscription extent $S_i$ changes, the list of intersections can be
recomputed in time $O\left( \min\{m, (k+1) \log m\}\right)$ using
$T_U$. Maintenance of both $T_U$ and $T_S$ does not affect the
asymptotic cost of~\ac{ITM}.

\section{Experimental Evaluation}\label{sec:experimental-evaluation}

The performance of a~\ac{DDM} service can be influenced by many
different factors, including: (\emph{i}) the computational cost of
the~\ac{DDM} matching algorithm; (\emph{ii}) the memory footprint;
(\emph{iii}) the communication overhead of the parallel/distributed
architecture where the simulation is executed and (\emph{iv}) the
cost of sending and discarding irrelevant events at the destination,
if any.

The communication overhead depends on the hardware platform over which
the simulation model is executed, and also on the implementation
details of the communication protocol used by the simulation
middleware. Therefore, factor (\emph{iii}) above is likely to equally
affect any~\ac{DDM} algorithm in the same way.

The cost of discarding irrelevant notifications applies only to
approximate matching algorithms, such as~\ac{GB} matching, that can
report spurious intersections (unless spurious intersections are
cleaned up at the sender side). The~\ac{BF}, \ac{SBM} and~\ac{ITM}
algorithms do not suffer from this problem since they never return
spurious intersections. Besides, in~\cite{Raczy2005}
and~\cite{6147978} the authors show that for relevant cases
the~\ac{SBM} algorithm has better performance than~\ac{GB} matching.
Therefore, in the performance evaluation study we focused on the exact
matching algorithms above, where only factors (\emph{i}) and
(\emph{ii}) should be considered.

\begin{table}[t]
\caption{\ac{DDM} algorithms considered in the experimental evaluation}\label{tab:alg-summary}
\centering\begin{tabular}{lll}
\toprule
\textbf{Algorithm} & \textbf{Computational Cost} & \textbf{Additional} \\
& & \textbf{Space} \\
\midrule
Brute Force & $O(nm)$ & none \\
Sort-Based & $O\left( (n+m) \log ( n+m ) + nm \right)$ & $O(n+m)$ \\
Interval Tree & $O\left( \min\{mn, (K+1) \log n\}\right)$ & $O(n)$ \\
\bottomrule
\end{tabular}
\end{table}

Table~\ref{tab:alg-summary} summarizes the computational and memory
requirements of the~\ac{DDM} algorithms considered in the experimental
evaluation. All costs are expressed in term of the number of
subscription extents $n$, update extents $m$, and total number of
intersections $K$. The ``Additional Space'' column specifies the
additional memory required by each algorithm, excluding the space
needed to maintain the lists of intervals, and excluding also the
space required by the intersection matrix. As discussed in the
previous sections, an extra $O(nm)$ space is required by all
algorithms to store the full intersection matrix.

It is important to observe that the asymptotic costs reported in
Table~\ref{tab:alg-summary} may have little significance when
evaluating the actual performance of the algorithms, since in practice
the constants hidden in the asymptotic notation may play a major role
and should not be ignored. For example, as already explained in
Section~\ref{sec:ddm-algorithms}, the weight of the term $nm$ in the
cost of~\ac{SBM} is likely very low since it is originated from simple
operations on bit vectors. For these reasons, we performed a set of
experimental evaluations whose outcome will be illustrated in this
section.

For better comparability of our results with those reported by other
research papers, we considered $d=1$ dimensions, and used the
methodology and parameters employed in~\cite{Raczy2005}. The first
parameter is the total number of extents $N$. We considered a total
number of extents in the range $[50 \times 10^3, 500 \times 10^3]$. In
all cases, $n = N/2$ are subscription extents and $m = N/2$ are update
extents. All extents are randomly placed on a segment of total length
$L=1 \times 10^6$. All extents have the same length $l$ that is
computed in order to obtain the desired overlapping degree $\alpha$,
defined as:

\begin{equation*}
\alpha=\frac{\sum \mbox{area of extents}}{\mbox{area of the routing space}} = \frac{N \times l}{L}
\end{equation*}

Therefore, for a given value of $\alpha$ and $N$, the length $l$ of
each segment can be computed as $l = \alpha L / N$. The overlapping
degree is an indirect measure of the total number of intersections
among subscription and update extents. While the cost of~\ac{BF}
and~\ac{SBM} is not affected by the number of intersections, this is not
the case for~\ac{ITM}. We considered the same values for $\alpha$ as
in~\cite{Raczy2005}, namely $\alpha \in \{0.01, 1, 100\}$.

The experimental evaluation has been performed on an Intel(R) Core(TM)
i7-2600 \@ 3.40 GHz CPU with 4 physical cores with Hyper-Threading
(HT) technology~\cite{HT}. The system has 16 GB of RAM and runs Ubuntu
11.04 (x86\_64 GNU/Linux, 2.6.38-16-generic \#67-Ubuntu SMP). HT works
by duplicating some parts of the processor except the main execution
units. From the point of view of the Operating System, each physical
processor core corresponds to two logical processors. Many studies
from Intel and others have shown that when HT is available, many
multi-threaded applications can have a performance boost in the range
from $16$ to $28\%$~\cite{HT}.  The algorithms have been implemented
in C and compiled with gcc version 4.5.2 using the \verb+-O3+
flag. The~\ac{SBM} algorithm has been implemented according to the
improved version described in~\cite[Section 4.2]{Raczy2005}, which is
more efficient than the basic version shown in
Algorithm~\ref{alg:sort-based}. The parallel versions of~\ac{BF}
and~\ac{ITM} have been obtained from the sequential implementation by
enabling OpenMP~\cite{openMP} directives in the code.

The performance metric of interest is the total (wall clock) execution
time needed to compute the intersection matrix; this time always
includes any preprocessing (e.g., the time required by~\ac{ITM} to
build the Interval Tree or the time needed by~\ac{SBM} to sort the
vector of endpoints). Each measure is the average of 30 independent
executions, in order to get statistically valid results. To foster the
reproducibility of our experiments, all the source code used in this
performance evaluation, and the raw data obtained in the experiments
execution, are freely available on the research group
website~\cite{pads} with a Free Software license.

\paragraph*{Sequential implementation}

\begin{figure*}[t]
\centering%
\subfigure[$\alpha=0.01$]{\includegraphics[width=.32\textwidth]{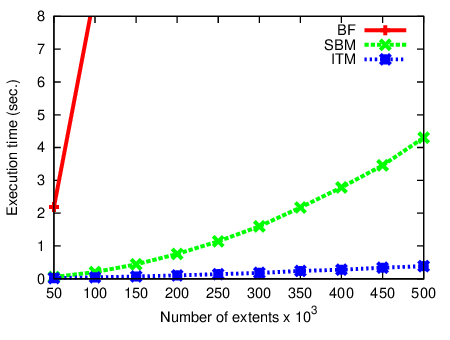}}
\subfigure[$\alpha=1$]{\includegraphics[width=.32\textwidth]{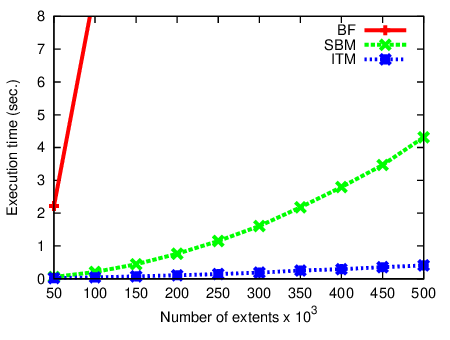}}
\subfigure[$\alpha=100$\label{fig:alfa100}]{\includegraphics[width=.32\textwidth]{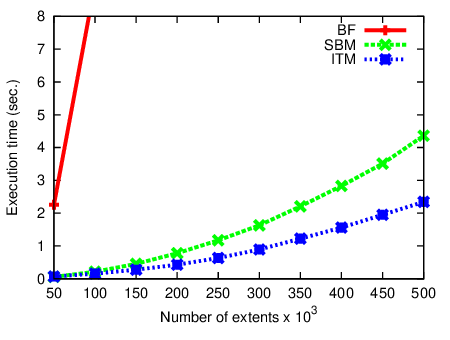}}
\caption{Execution time in seconds as a function of the number $N$ of
  extents (lower is better). The vertical linear scale, as opposed to
  logarithmic scale, has been defined to allow easier comparison
  between~\ac{SBM} and~\ac{ITM}}\label{fig:sequential_comparison}
\end{figure*}

We start by comparing the performance of the sequential
implementations of~\ac{BF}, \ac{SBM} and~\ac{ITM}.
Figure~\ref{fig:sequential_comparison} shows the total execution time
of each algorithm as a function of the number of extents $N$, with
low, medium and high overlapping degrees $\alpha$.

The vertical scale has been set to allow an easier comparison
of~\ac{SBM} and~\ac{ITM}; since~\acl{BF} is the slower algorithm, its
execution time goes quickly out of scale. We see that~\ac{ITM} is
faster than~\ac{SBM}, but the gap between them tend to close as the
overlapping degree $\alpha$ grows. We also observe that the execution
time of~\ac{SBM} is unaffected by the value of $\alpha$, which is
expected by observing in Table~\ref{tab:alg-summary} that its cost
does not depend on $K$.

\paragraph*{Parallel implementation}

We now study the performance of a parallel implementation of~\ac{ITM}.
As stated in Section~\ref{sec:parallel-ddm}, \ac{ITM} can be trivially
parallelized using a multi-thread implementation in which each
thread is assigned a subset of the queries. Parallel versions
of~\ac{ITM} and~\ac{BF} have been obtained by enabling OpenMP
directives in the source code.  To the best of our knowledge no
parallel versions of~\ac{SBM} have been proposed, and as explained in
Section~\ref{sec:ddm-algorithms} the sequential step of~\ac{SBM} is
affected by a loop-carried dependency which can not be easily avoided.

\begin{figure}[t]
\centering%
\includegraphics[width=.9\columnwidth]{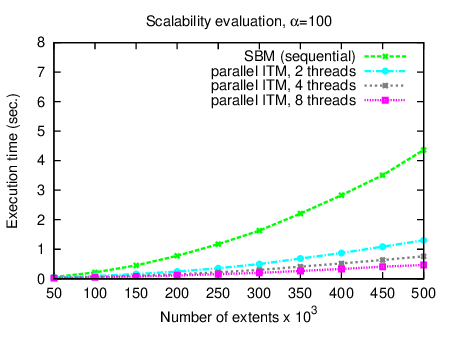}
\caption{Execution time of~\ac{SBM} (sequential) and~\ac{ITM}
  (parallel), $\alpha=100$, lower is
  better}\label{fig:parallel_comparison_wct}
\end{figure}

Figure~\ref{fig:parallel_comparison_wct} shows the execution time of
the sequential~\ac{SBM} with three configurations of~\ac{ITM} in which
a different number of concurrent threads is used. We consider the case
in which $\alpha=100$, corresponding the the scenario depicted in
Figure~\ref{fig:alfa100}. As expected, the parallel~\ac{ITM} can
exploit multiple processor cores to increase the gap from~\ac{SBM}.
The practical effect is that~\ac{ITM} remains competitive for larger
number of intersections.

\begin{figure}[t]
\centering%
\includegraphics[width=.9\columnwidth]{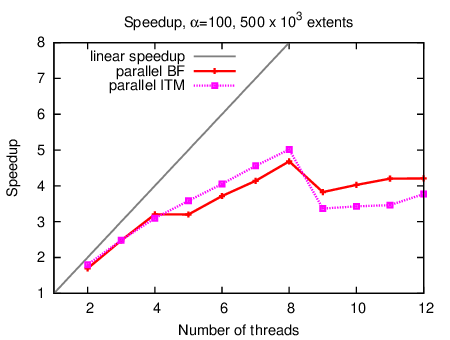}
\caption{Speedup of parallel~\ac{BF} and parallel~\ac{ITM}, $N=500 \times 10^3$ extents, $\alpha=100$, higher is better}\label{fig:parallel_comparison_speedup}
\end{figure}

To understand the scalability of the parallel version of~\ac{ITM} we
compute the speedup $S_p$ as a function of the number $p$ of execution
threads, where $S_p$ is defined as the ratio of the execution time of
the sequential implementation and the execution time with $p$ threads.

Figure~\ref{fig:parallel_comparison_speedup} shows the speedup of
parallel~\ac{BF} and parallel~\ac{ITM}. Despite its inefficiency in
terms of wall-clock time, \ac{BF} has been considered here because it
is so easily parallelizable that it provides baseline values for the
speedup. Interestingly, \ac{ITM} scales better than~\ac{BF}; this is
probably due to the improved locality achieved by the Interval Tree,
since the intervals stored by the nodes near the root are likely kept
in cache. Figure~\ref{fig:parallel_comparison_speedup} shows also the
effect of HT: when the number of threads $p$ is the in the range $1,
\ldots, 4$ each execution thread is allocated on a dedicated physical
core. When the number of threads exceeds the number of physical cores,
then multiple threads are allocated to the same core. As said before,
HT does provide a performance boost, but unfortunately not comparable
to that provided by an actual physical core with independent execution
units. The speedup drops when the number of threads exceeds the number
of logical cores.
 
\begin{figure}[t]
\centering\includegraphics[width=\columnwidth]{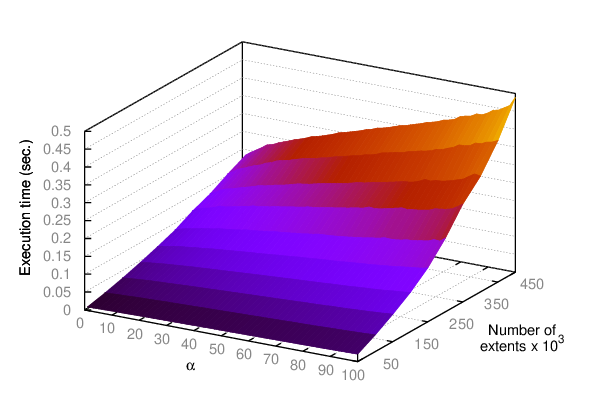}
\caption{Execution time of parallel~\ac{ITM}, 8 execution threads, lower is better}\label{fig:parallel_ITM_scalability_evaluation}
\end{figure}

Finally, Figure~\ref{fig:parallel_ITM_scalability_evaluation}
summarizes the execution time of parallel~\ac{ITM} over a range of
overlapping degrees ($0 < \alpha \leq 100$) and for an increasing
number of extents (from $50 \times 10^3$ to $500 \times 10^3$). The
plot supports the fact that the running time of~\ac{ITM} depends on
both the number of matches $K$ and the input size $N$; in our
experiments the value of $K$ is directly correlated with the overlap
factor $\alpha$, hence the shape of the graph.

\section{Conclusions and Future Works}\label{sec:conclusions}

In this paper we described~\ac{ITM}, a parallel algorithm based on
Interval Trees that can be used to solve the $d$-rectangle
intersection problem for~\ac{DDM}. \ac{ITM} uses an augmented AVL tree
data structure to store a set of intervals, allowing fast intersection
queries. \ac{ITM} is of practical interest since it can be implemented
quite easily; a prototype has been built and is available
at~\cite{pads}.  Both the sequential and the parallel implementation
of~\ac{ITM} has been evaluated experimentally; the results show that
the sequential implementation of~\ac{ITM} compares favorably
with~\acl{SBM}, the current best solution to the~\ac{DDM} matching
problem. The parallel version of~\ac{ITM} shows good scalability,
achieving a $\tilde 5$ speedup on a four core, hyperthreaded Intel i7
processor.

We are currently extending the~\ac{ITM} prototype as described in
Section~\ref{sec:parallel-ddm} to solve the dynamic~\ac{DDM} matching
problem, where extents can be moved or resized
dynamically. Furthermore, we are including support for~\ac{ITM} in the
GAIA/ART\`IS parallel simulation middleware~\cite{gda-dsrt-2004}. This
will allow us to test~\ac{ITM} in real simulation models, to further
assess its performance.

\section*{Acknowledgment}

The authors would like to thank Piero Fariselli for kindly providing
the server on which the experimental evaluation described in
Section~\ref{sec:experimental-evaluation} has been done.

\section*{Notation}

\begin{center}
\begin{tabular}{rcl}
$\mathbf{S}$ & := & Subscription set $\mathbf{S} = \{S_1, \ldots, S_n\}$ \\
$\mathbf{U}$ & := & Update set $\mathbf{U} = \{U_1, \ldots, U_m\}$ \\
$n$ & := & Number of subscription extents \\
$m$ & := & Number of update extents\\
$N$ & := & Total number of subscription and update extents\\
$\mathbf{M}$ & := & $n \times m$ intersection matrix\\
$K$ & := & Number of intersections, $K \leq nm$ \\
$\alpha$ & := & Overlapping degree\\
\end{tabular}
\end{center}


\end{document}